# Correlating spin and optical properties of quantum emitters in hBN

Nika Teran[1,2], Benjamin Whitefield[1,2], Nicholas Sloane[1,2], Kenji Watanabe[3], Takashi Taniguchi[4], Igor Aharonovich[1,2] and Mehran Kianinia[*,1,2]

[1] *School of Mathematical and Physical Sciences, University of Technology Sydney, Ultimo, New South Wales 2007, Australia*
[2] *ARC Centre of Excellence for Transformative Meta-Optical Systems, University of Technology Sydney, Ultimo, New South Wales 2007, Australia*
[3] *Research Center for Electronic and Optical Materials, National Institute for Materials Science, 1-1 Namiki, Tsukuba 305-0044, Japan 7 Research Center for Materials Nanoarchitectonics, National [4]Institute for Materials Science, 1-1 Namiki, Tsukuba 305-0044, Japan305-0044, Japan*These authors contributed equally*

[*] To whom correspondence should be addressed: mehran.kianinia@uts.edu.au

## Abstract

Optically addressable spin defects in hexagonal boron nitride (hBN) are well suited for near-surface quantum sensing. They offer bright, wavelength-tunable single-photon emission with high optically detected magnetic resonance (ODMR) contrast at room temperature. Here we controllably synthesise a high density of spin-complex defects in carbon-doped hBN flakes. We find that the zero-field splitting parameter *D* is directly correlated with the zero-phonon line of an emitter, while the ODMR contrast shows no such correlation. We further analyse an individual narrowband defect showing 68% ODMR contrast and enhance its photon collection by ~40% using a solid immersion lens. By advancing both the practical synthesis of the spin complexes and the understanding of their microscopic origin, our results move us toward the deterministic creation of ODMR-active quantum emitters.

## Introduction

Solid-state quantum emitters with optically-addressable spin systems underpin a range of quantum technologies, from nanoscale magnetometry to quantum networks.[1–11] Beyond three-dimensional systems operating at room temperature like the well-explored nitrogen-vacancy (NV) centre in diamond[12,13] and defects in silicon carbide,[14,15] colour centres in two-dimensional hexagonal boron nitride (hBN) have emerged as an attractive alternative. The platform allows manipulation of spin in the photoluminescent defects within nanometres of a surface, while also being easily integrated with other van der Waals materials and existing semiconductor technologies.[16–23]

Among optically addressable spin defects in hBN, the negatively charged boron vacancy has been studied most extensively.[22,24–29] However, it has only been observed in ensembles so far, imposing limitations on its quantum coherent applications. Recently, a family of bright, narrowband single-photon emitters was shown to exhibit optically detected magnetic resonance (ODMR) across a broad range of wavelengths spanning the visible and near-infrared spectrum.[30–32] In contrast to the NV centre with a dim zero phonon line (ZPL) fixed at 637 nm and broad emission from the phonon sideband, these defects allow quantum light emission to be matched to a given target system. Moreover, the reported ODMR contrasts of these hBN emitters reach up to 200%.[31,33,34]

The spin complex model involving charge transfer between two nearby defects is currently the most widely accepted interpretation of ODMR activity in these emitters.[31,33–39] The model is characterised by two regimes illustrated in Figure 1b. The ZPL (Figure 1c) corresponds to the spin-allowed singlet (S = 0) transition in the strongly coupled regime (shaded orange in Figure 1b) when both electrons reside on defect A. This configuration also gives rise to the S = 1 transitions ($\nu_+$ and $\nu_-$ shaded orange in Figure 1d). The zero-field splitting (ZFS) parameters *D* and *E*, which characterise these transitions, are therefore also intrinsic to defect A. In the weakly coupled regime (shaded blue in Figure 1b), one electron resides on defect A and the other on a remote defect B following charge transfer. The spins are weakly coupled due to the separation of at least 1 nm.[35] This interaction gives rise to an effective S = ½ transition at the Zeeman frequency $\nu_{½}$ (shaded blue in Figure 1d). This transition is therefore governed by the separation between the two defects rather than defect A alone. The full description of the spin complex model is given in the Supplementary Information Section I.

Here, we study ODMR-active emitters generated in carbon-doped hBN flakes by oxygen annealing (Figure 1a). We first optimise the annealing process for compatibility with thin flakes that can be used in integrated photonic devices. We realise high emitter density, crucial for practical use as only a portion of all emitters show ODMR activity. Secondly, we report the observed trends between optical and spin parameters of individual emitters, further supporting the proposed spin complex model. Finally, we characterise an individual emitter, combining narrowband quantum emission with high-contrast room-temperature spin readout and enhance its photon collection with a solid immersion lens.

## Results and discussion

Our goal was to obtain high emitter density with significant ODMR-active emitter yield, while preserving flake integrity. We conducted a time and temperature sweep based on the single-step thermal annealing approach of Whitefield *et al.*[31] Exfoliated carbon-doped hBN flakes were annealed in an oxygen atmosphere for a time period ranging from 1 to 4 hours. Shorter annealing times (1 and 2 hours) yielded negligible amounts of emitters. As the time increased to 4 hours, we observed a substantial increase in density of emitters (Figure 2a). We therefore infer that 4 hours is the optimal time parameter for our method.
Thickness of the flakes was measured using atomic force microscopy (AFM) before and after annealing at different temperatures (Supplementary Information Section II). As shown in Figure 2b, we observed minimal etching at 900ºC, however the thickness of flakes annealed at 1000ºC reduced by 32 nm on average. Decomposition of hBN above 850ºC was previously reported even in an inert environment, however moderate etching at higher temperatures was found to yield higher concentrations of stable SPEs. Ahmed *et al.*[40] reported that surface contaminants act as nucleation sites for the etching process. Hence, the flakes were also cleaned in isopropyl alcohol (IPA) prior to annealing to remove any residual debris from the exfoliation process. Nevertheless, etching of the IPA cleaned sample at 900ºC was not found to be any different from that of the sample that was cleaned only in UV-ozone (Figure 2b).

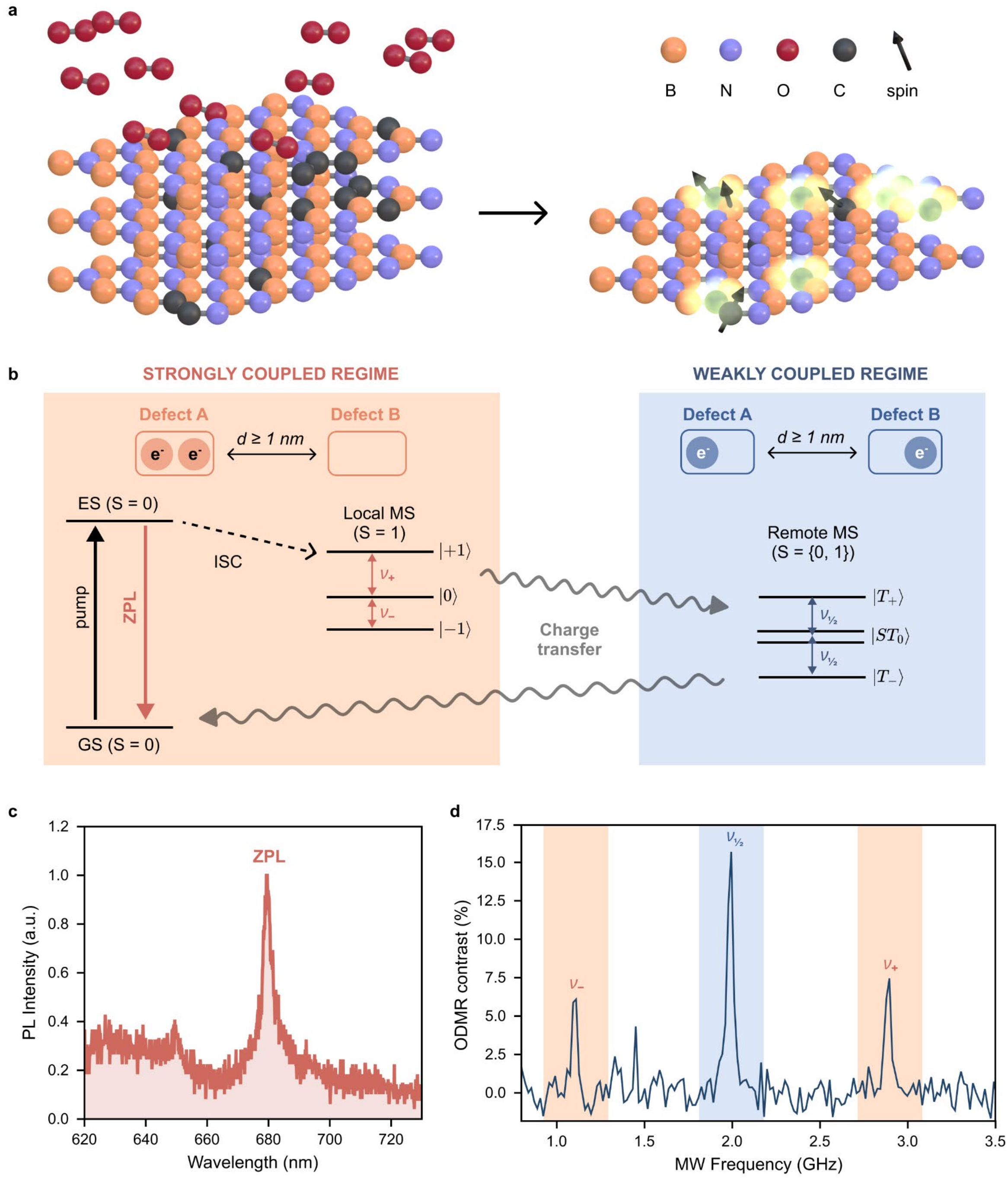

**Figure 1. Generation of ODMR-active quantum emitters in hBN flakes described by the spin complex model. (a)** Schematic of annealing a carbon-doped hBN flake in an oxygen atmosphere (left) to create ODMR-active emitters where etching of the top hBN layer occurs. **(b)** Simplified energy level structure of the proposed spin-complex model. The orange panel on the left represents both electrons residing on defect A, while the blue panel on the right shows the regime after charge transfer, when one electron remains on defect A and the other is localised on defect B. The distance between defects is denoted by *d*. The radiative transition from the excited state (ES) to ground state (GS) is shown by the solid orange arrow, while the dashed arrows depict non-radiative transitions through the metastable states (MS). **(c)** Example photoluminescence (PL) spectrum with a zero-phonon line (ZPL) corresponding to the radiative transition described in **(b)**. **(d)** ODMR spectrum measured from the emitter.

We conclude that annealing carbon-doped hBN flakes at 900ºC for 4 hours is optimal for obtaining high emitter density while retaining flake thickness. Using these parameters, we produced emitters with narrowband zero phonon lines (ZPL) spanning the characteristic 560-750 nm range in agreement with previous studies.[31,41,42] Reduced etching allows this method to be used for flakes of less than 10 nm in thickness while even thinner flakes, possibly monolayers, could be used with further optimisation.

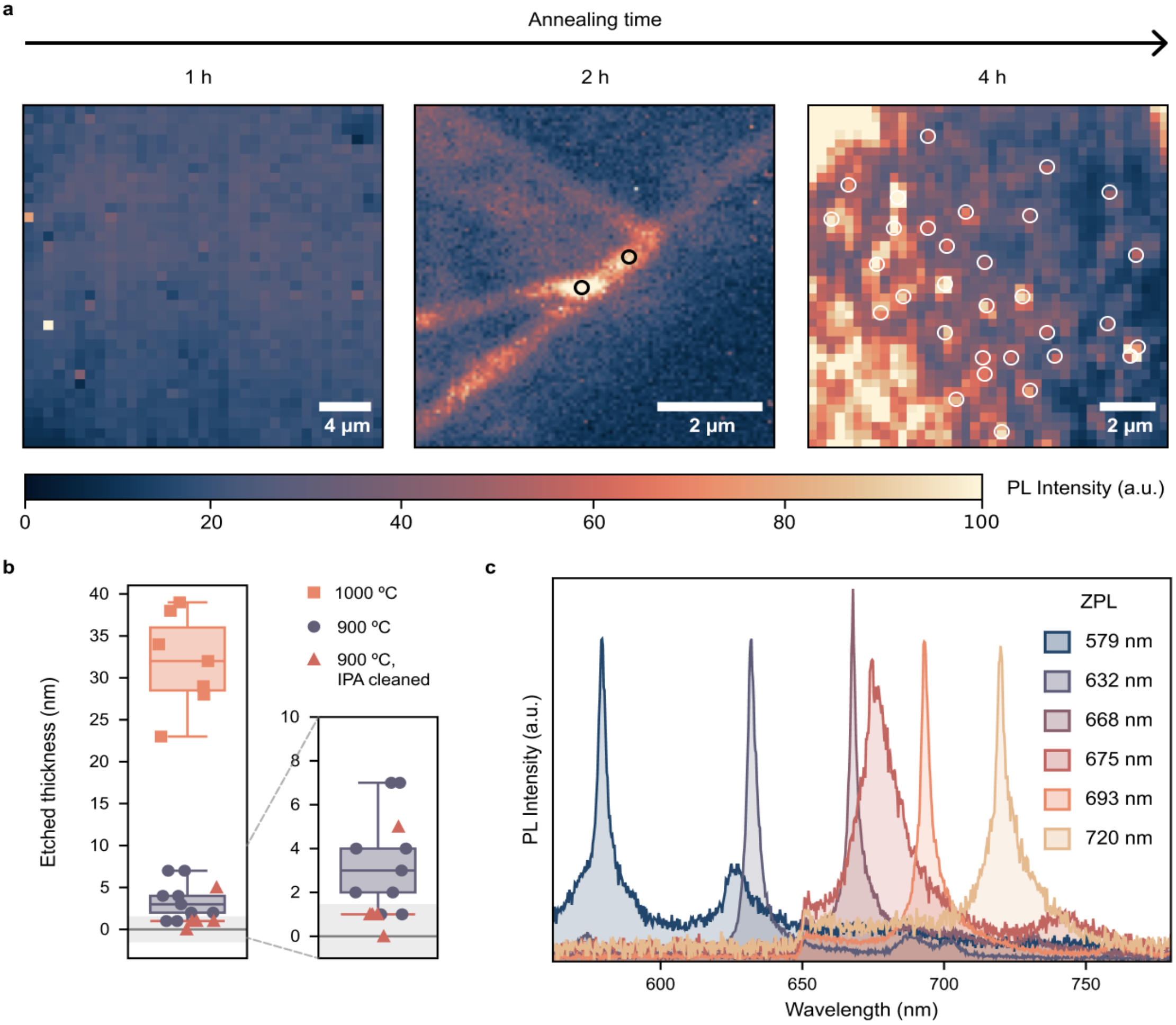

**Figure 2. Oxygen annealing optimisation for high density of visible emitters and retained integrity of exfoliated hBN flakes. (a)** Confocal PL scans of hBN flakes annealed at 1000ºC for 1, 2, and 4 hours. Emitters are circled in black or white. **(b)** Box plots summarising the amount of hBN etched during the annealing process at 1000ºC (orange squares) and 900ºC (blue circles). Zoomed-in axes show that no significant difference was observed between flakes cleaned only in UV-ozone before annealing, and flakes additionally cleaned in IPA (triangles). **(c)** Emission spectra of emitters resulting from a 4h anneal at 900ºC (Figure S3b).

Flake A (Figure 3a), which possesses a high density of emitters, was chosen for the statistical analysis of the generated emitters. Out of 181 emitters, 25 (14%) showed $\nu_{½}$ ODMR contrast (Figure 3b). We could accurately determine the ZPLs of 12 ODMR-active emitters, and the $\nu_{+}$ and $\nu_{-}$ transitions from 9 of this subset. To gain insight into emitter characteristics, pairwise correlations of their properties were quantified using Pearson $r$, Spearman $\rho$, and their respective two-sided $p$-values for sample sizes $n$ (Supplementary Information Section III). We assessed the ZPL energy, ZPL full width half maxima (FWHM), the axial ZFS parameter $D$, and ODMR contrast of the $\nu_{½}$, $\nu_{-}$, and $\nu_{+}$ transitions. The transverse ZFS parameter $E$ could not be reliably extracted for all emitters (Supplementary Information Section Ia). We therefore do not include it in our analysis. Given the small sample size $n$, Spearman $\rho$, which is less sensitive to outliers, is reported in the main text rather than Pearson $r$, which is sensitive to outliers and assumes a linear relationship (reported in Tables S1-4).

Figure 3e, f show the Spearman $\rho$ in the form of pairwise correlation matrices where red signifies positive correlation ($\rho \to +1$) and blue negative correlation ($\rho \to -1$), while white/neutral indicates no correlation ($\rho \to 0$). The size of the circles corresponds to the strength of the correlation with larger discs indicating lower p-value and therefore higher significance.

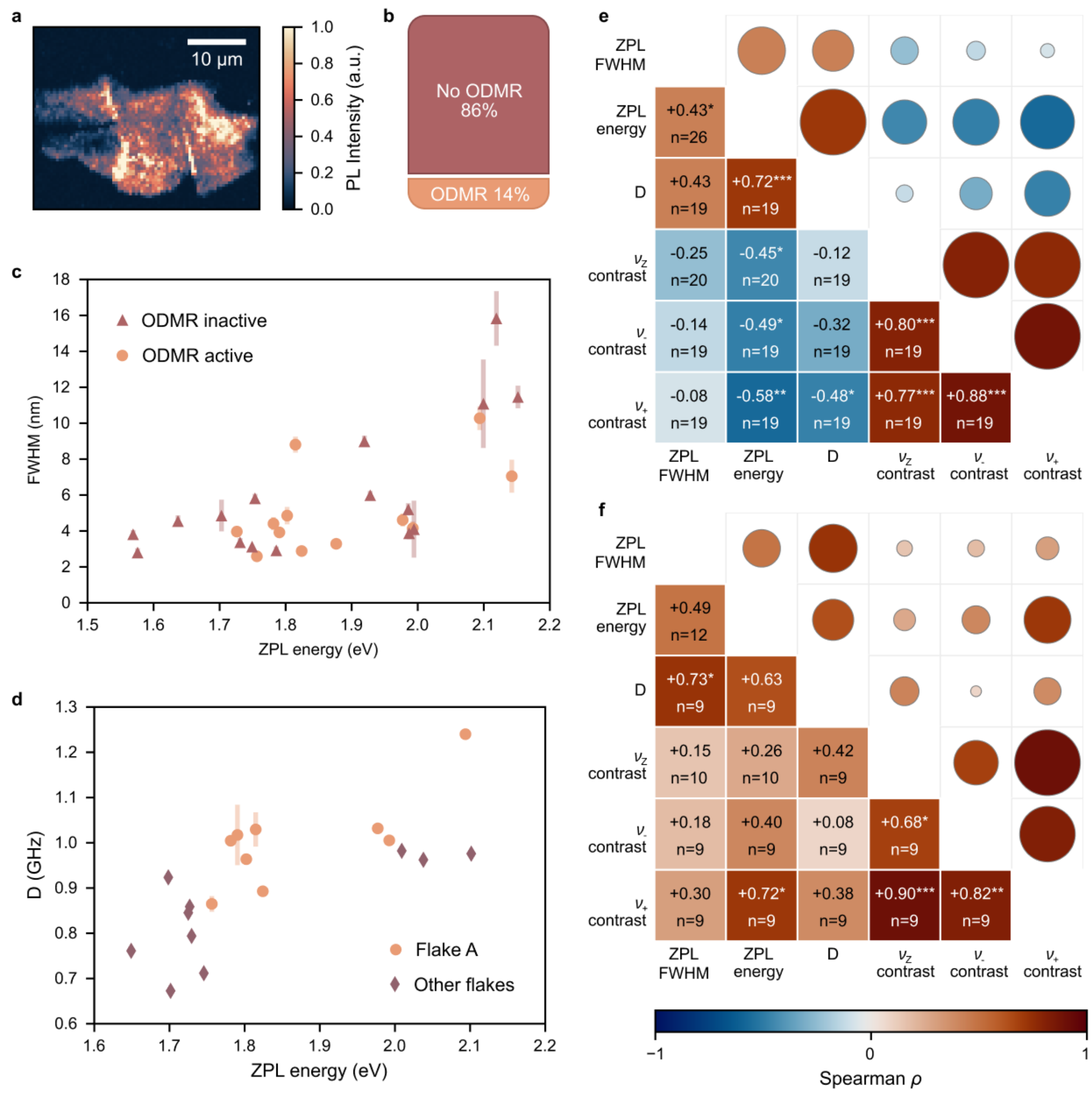

**Figure 3. Intra- and inter-flake emitter statistics for ODMR-active and inactive emitters. (a)** Confocal PL scan of Flake A. **(b)** Visualisation of proportion of ODMR-inactive (top) and active (bottom) emitters in Flake A. **(c)** Scatter plot of ZPL FWHM versus ZPL energy of emitters in Flake A coded by ODMR-active (circles) and inactive (triangles). **(d)** Scatter plot of ZFS parameter *D* versus ZPL energy of ODMR-active emitters in Flake A (circles) and other flakes (diamonds). **(e)**, **(f)** Correlation matrices of Spearman *ρ* of ZPL FWHM, ZPL energy, *D*, and ODMR contrast of the $\nu_{½}$, $\nu_{-}$, and $\nu_{+}$ transitions grouped by **(e)** pooled emitters and **(f)** only emitters in Flake A. Each pair is shown twice. In the lower left, colours of the boxes represent the Spearman *ρ* value printed above the corresponding number of emitters *n* used for its computation. The significance of the values according to the *p*-value is denoted by asterisks (* $p < 0.05$, ** $p < 0.01$, *** $p < 0.001$). In the top right, the discs encode *ρ* in identical colour scale to the boxes and have areas proportional to $-log_{10}(p)$ – larger discs indicate higher statistical significance, saturating at $p \leq 0.001$. Colour scale: common for **(e)** and **(f)**; diverging from dark blue (strong negative correlation), through white/neutral (no correlation) to dark red (strong positive correlation).

First, the optical properties of both ODMR-active and inactive emitters within Flake A were tested. The ZPL energy and linewidth were found to be significantly correlated ($p < 0.001$).

Correlated microscopy and resonant spectroscopy studies have established that hBN emitters in this range comprise multiple distinct defect species rather than a single strain-tuned defect, and that local strain is neither required to activate emitters nor sufficient to account for the ZPL spread.[42,43] We therefore attribute the correlation to systematic differences among defect configurations rather than to strain acting on a common defect.

The proposed structure of defect A is a donor-acceptor pair (DAP), most likely consisting of carbon substitutional defects (e.g. $C_B$ and $C_N$) separated by a multiple of the B-N bond length (~1.5 Å)[44]. Since the DAP is separated merely by a few Å, it behaves as a single bright emitter with a narrow ZPL, rather than as a long-range DAP recombination.[35] The existence of the ZPL energy–FWHM trend supports the hypothesis suggesting a family of defects, i.e. DAPs with several possible distances between the donor and the acceptor, which serve as the optically-active species (defect A) of the spin complex. In previous theoretical calculations[31,35], the increase in internal DAP distance was predicted to decrease the ZPL energy. Combined with our results, this indicates that when the donor and acceptor separation increases, the ZPL FWHM decreases. The room-temperature FWHM reflects a combination of electron-phonon dephasing and spectral diffusion due to charge noise,[41] and our measurements do not separately address these contributions. Therefore, the reported empirical energy-linewidth correlation cannot be attributed to a specific mechanism. Nevertheless, based on the assumption that defect A can be a DAP, we suggest that more compact defect complexes, which emit at higher energy, experience more internal fluctuations due to vibrations of the constituent atoms. The constant thermal vibrations can therefore smear the emission over a wider range of energies, broadening the ZPL. The correlation was present in both the ODMR-active ($p = 0.106$ – borderline significant) and inactive ($p = 0.004$) emitter subsets. In Figure 3c, the overlapping scatter plots of both groups show they trend in the upward direction along a similar baseline, suggesting that optical properties of these emitters are not indicative of ODMR activity.

Next, we investigate pairwise correlations of ODMR-active emitters exclusively, to elucidate any trends in ODMR activity. The correlation matrix in Figure 3e includes emitters across different flakes, while Figure 3f describes ODMR-active emitters exclusively from Flake A. First, we turn our attention to correlations between ODMR contrasts of $\nu_{½}$, $\nu_{-}$, and $\nu_{+}$ transitions. Since the contrasts are all extracted from the same ODMR measurement, we expect near-unity correlations for all of them, regardless of the emitter. The pooled data in Figure 3e confirms that. Any deviations from $\rho = +1$ are attributed to frequency-dependent microwave delivery. Namely, transitions lie at considerably different frequencies (~1, 2, and 3 GHz) where the amplifier gain and waveguide response differ. Data from Flake A (Figure 3f) exhibits a slight change in those correlations, which is most likely due to the smaller sample size. Therefore, we note that other correlations computed with $n = 9\text{-}12$ should be interpreted carefully and conclude that ODMR contrasts of $\nu_{½}$, $\nu_{-}$, and $\nu_{+}$ transitions either increase or decrease together.

A significant correlation is shown between $D$ and ZPL energy for all ODMR-active emitters ($p < 0.001$; $n = 19$). It becomes borderline significant for emitters in Flake A ($p = 0.067$; $n = 9$), possibly due to small $n$. The increasing trend of $D$ with ZPL energy is evident in the scatter plot in Figure 3d. This correlation between the ZFS parameter and the optical properties of the emitter is consistent with the spin complex model where the two electrons have a common origin of defect A. Furthermore, the magnitude of the spin-spin coupling $D \sim 1$ GHz found here is of the same order as previously calculated *1.4 GHz* and *1.8 GHz* for DAP internal separation of 2 and $\sqrt{7}$ B-N bonds, respectively[31], supporting the identification of defect A as a DAP with spins in proximity of a few Å.

Finally, we comment on the correlations between ODMR contrast and other parameters. All of them negatively correlate with ZPL energy for pooled data, however the correlation reverses sign within Flake A. Unlike ZPL energy, ZPL FWHM and $D$, ODMR contrast is not solely an intrinsic property of the emitter. It is the fractional PL change that depends on the optical excitation power, photon purity, the microwave drive field and its alignment with the emitter in addition to the underlying spin-dependent charge-transfer dynamics. These conditions are homogenous only within a single measurement and differ substantially between flakes and measurements. Pooling the values across flakes consequently introduces differences unrelated to the optical properties of emitters, and generates apparent correlations shown in Figure 3e. We therefore cannot draw meaningful conclusions from the pooled correlations.

The variation in measurement conditions is minimised within a flake, which allows us to derive some important conclusions based on the correlations shown in Figure 3f. No contrast shows a significant correlation to the optical properties. The single exception, ZPL – $\nu_+$ contrast, is not reproduced for the other two contrasts despite their mutual correlation, so we cannot interpret it as indicative in this dataset. The lack of correlation between ODMR contrast and optical properties is consistent with the spin complex model, where the ODMR-inducing mechanism requires coupling of defect A to defect B that is at least 1 nm away. The A to B distance varies independently of the DAP structure and therefore the optical properties of the emitter. Consequently, the ODMR contrast does not appear to be related to the ZPL properties either.

The preceding analysis establishes how the spin and optical parameters relate to one another across the family of emitters. To illustrate the absolute performance of an individual emitter, we perform a detailed photon-statistical characterisation of an isolated defect (Figure 4a). Its photoluminescence (PL) spectrum shows a ZPL at *718 nm* (*1.73 eV*) with a Lorentzian linewidth of *5.61 nm* and an excited state lifetime of *6.35 ns* at room temperature (Figure 4b). The second-order autocorrelation, measured without background correction, confirms single photon emission with $g^{(2)}(0) = 0.33$ (Figure 4c). All of these parameters lie within the characteristic values for emitters described by the spin complex model in this work and in previous studies,[31,33,39] with the lifetime being longer than most previously reported values.

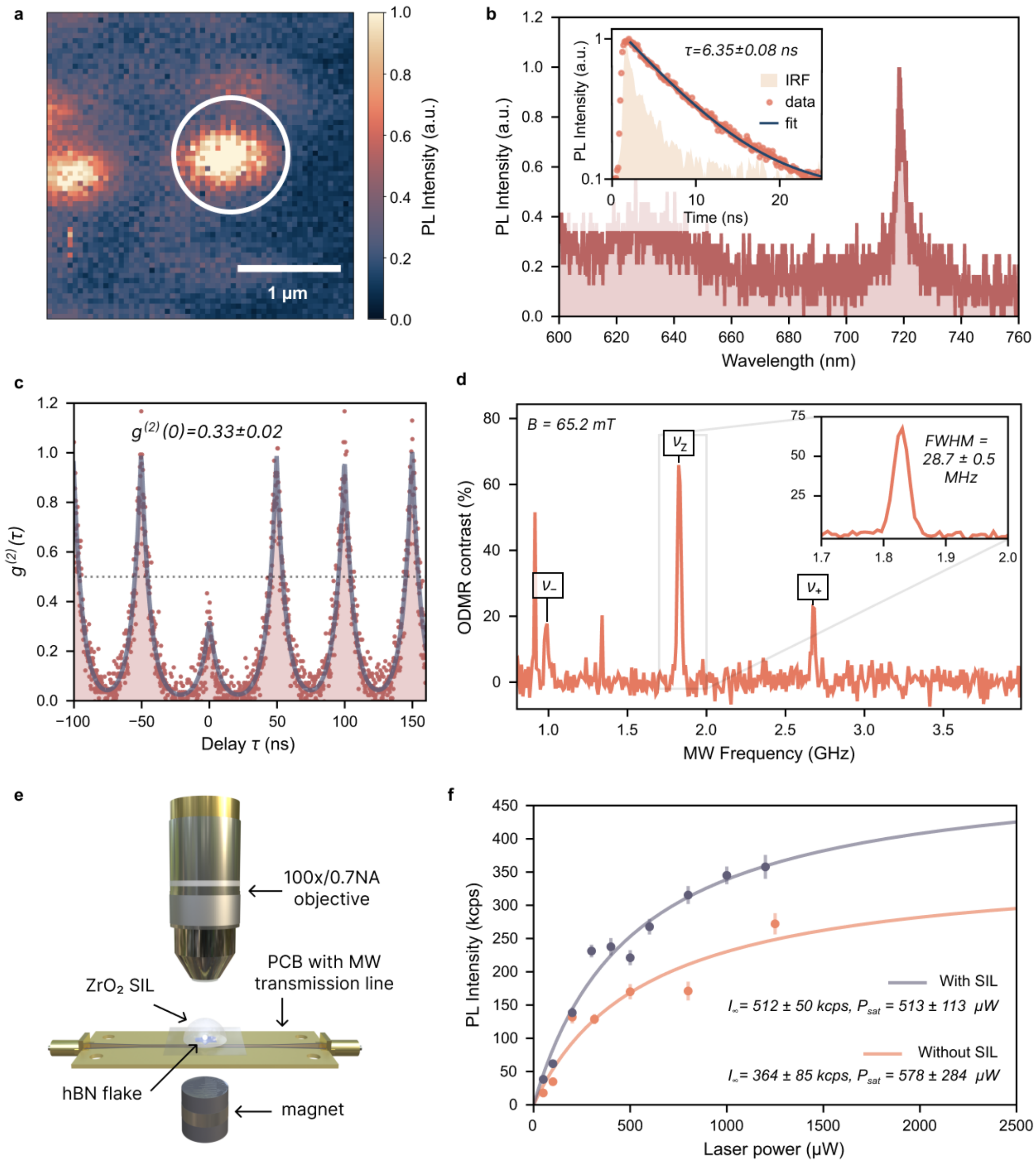

**Figure 4. Characteristics and enhancement of an ODMR-active single photon emitter. (a)** Confocal PL scan of the emitter with a *720 nm* band-pass filter, circled in white. **(b)** PL emission spectrum. Inset: PL intensity decay after pulse, fitted by a single exponential function to obtain emitter's optical lifetime and corresponding instrument response function (IRF). **(c)** *20 MHz* pulsed second order autocorrelation function, $g^{(2)}(\tau)$, with a dip $g^{(2)}(0) < 0.5$ indicating single photon emission without background correction. The dashed horizontal line at $g^{(2)}(0) = 0.5$ is a guide to the eye. **(d)** ODMR contrast spectrum with annotated $\nu_{1/2}$, $\nu_-$, and $\nu_+$ transitions measured at an applied magnetic field strength $B = 65.2\ mT$. The unlabeled peaks arise from the second harmonics of the $\nu_{1/2}$ and $\nu_+$ transitions from left to right. Inset: zoomed-in peak with annotated FWHM. **(e)** Schematic of the sample used to enhance emission with a ⌀ 2 mm zirconia solid-immersion lens (SIL). The SIL sits on top of an hBN flake that was transferred onto a glass coverslip. The flake is positioned over a coplanar waveguide on a printed circuit board (PCB). The magnet provides the applied field required for spin complex ODMR measurements. Both excitation and collection go through a 100× magnification 0.7 numerical aperture

(NA) objective. **(f)** Optical power saturation curves of the emitter measured with and without the SIL alongside annotated fit parameters.

Furthermore, the emitter's ODMR spectrum (Figure 4d) shows the three spin-complex characteristic resonances at $\nu_{-}$ = *0.99 GHz*, $\nu_{½}$ = *1.83 GHz*, and $\nu_{+}$ = *2.68 GHz*. We use them to determine the magnitude of the external magnetic field (*65.2 mT*) from the $\nu_{½}$ and the ZFS parameter *D* (*0.845 GHz*) using $\nu_{-}$ and $\nu_{+}$ values. Its $\nu_{½}$ contrast reaches *68%*, the highest observed in this study, with a FWHM of *28.7 MHz*. We calculate the DC magnetic field sensitivity of the emitter to be *2.7* $\mu T/Hz^{1/2}$ (Supplementary Information Section IV), which is comparable to that of single NV centres.[45]

To improve photon collection and further improve the sensitivity, we placed a ⌀ 2 mm zirconia solid-immersion lens (SIL) on top of the flake. Saturation measurements showed ~ 40% enhancement with the saturation power unchanged within uncertainty (Figure 4f), which indicates that the SIL improved photon collection without altering the photophysics of the emitter. Additionally, we take advantage of the robustness of hBN flakes and stability of these emitters, as we transfer the flake onto a microwave-transparent substrate. This enables the maintenance of the flake's position relative to the microwave source with or without the use of a SIL (Figure 4e).

## Conclusion

To conclude, we found the optimal conditions for realising high emitter density while retaining flake thickness using oxygen annealing of carbon-doped hBN flakes at 900 ºC for 4 hours. This method can be further explored for integrated devices and near-surface applications. We show that 14% of the generated emitters within a single flake exhibit ODMR activity. Across this population, we find that ODMR activity cannot be predicted from the ZPL properties. Namely, both the ODMR-active and inactive emitters follow the same trend where ZPL linewidth increases with ZPL energy. Instead, the ZFS parameter *D* was found to correlate with the optical properties of the emitter, whereas the ODMR contrasts do not. This leads us to conclude that defect A carries the optical emission and ZFS properties, while the ODMR depends on a different regime, namely the charge transfer to a remote defect B. Together with the magnitude of *D* ~1 GHz, the findings support identification of defect A as a compact donor-acceptor pair. Our population-level test through correlation of spin and optical properties is therefore complementary to previous studies on individual defects in the identification of the microscopic structure of the visible-range ODMR-active defects in hBN. Critically, our results provide convincing evidence in support of the proposed spin complex model. Finally, we characterised an individual emitter combining characteristic narrowband single-photon emission with high ODMR contrast at room temperature. We demonstrated a ~40% enhancement of the photon collection using a SIL. Together, these results position hBN single-photon emitters as a promising room-temperature wavelength-tunable platform for quantum sensing and integrated photonics, while the statistical analysis paves the way for refining the spin complex defect model and ultimately guiding the deterministic creation of ODMR-active emitters.

## Methods

**Sample preparation**: Carbon-doped hBN crystals[46] were mechanically exfoliated using scotch tape. The flakes were transferred onto a $SiO_2$/Si substrate using Gel-Pak at 60°C. The samples were then annealed in a quartz tube of the Lindberg Blue 3000 Furnace in vacuum (~ 30 mTorr) with constant oxygen flow at 1000 sccm. Oxygen-annealed flakes were dry-transferred onto a glass coverslip using a polydimethylsiloxane (PDMS) stamp and polyvinyl alcohol (PVA). PVA was dissolved in water and the sample cleaned in UV-ozone to remove any polymer residue. The coverslip with the hBN flakes was then attached onto a printed circuit board (PCB) so that the flakes were positioned directly over the MW transmission line. A thin glass coverslip was chosen as the optimal substrate to achieve closeness to the MW source as glass transmits it, while omitting the conventional wire over sample which can reduce optical visibility of emitters if too close.

**Optical measurements**: Optical measurements were carried out using a reflection-based home-built scanning confocal microscope with 532 nm continuous wave laser and a 0.7 NA 100 x objective (Olympus) with a 568 nm long-pass filter to filter out laser in collection with the samples mounted on a piezo stage. The light was collected with avalanche photodiodes (APD) (Excelitas). The correlation measurements were done using a 512 nm pulsed laser with a 20 MHz repetition rate. The collected photon counts were correlated to the laser pulses using a coincidence counter module (PicoQuant).

**Optically detected magnetic resonance (ODMR)**: A static magnetic field was applied using a permanent NdFeB magnet (grade N35) positioned perpendicular to the sample surface to lift the spin-state degeneracy. Microwave excitation was provided by a radio-frequency signal generator (AnaPico APSIN 4010), with the output delivered to the sample via the PCB. The microwave signal was amplified using a high-power microwave amplifier (Mini-Circuits ZHL-16W-43-S+). The microwave frequency was swept within the 0.6 – 4.0 GHz resonance range with each step consisting of 1 ms RF signal followed by a 1 ms off time to allow for a reference measurement, while monitoring the photoluminescence signal. For each step, the ODMR contrast was calculated using:

$$Contrast\ (\%) = \frac{Signal - Reference}{Reference} \times 100 \quad (1)$$

**Supporting information:** Description of the spin complex model, Calculation of ZFS parameters, Flake thickness and emitters density measurements, statistical analysis of spectral and ODMR measurements and sensitivity calculation.

## Acknowledgments

The authors acknowledge financial support from the Australian Research Council (CE200100010, FT220100053, and DP250100973). K.W. and T.T. acknowledge support from the Japan Society for Promotion of Science (JSPS) KAKENHI (grant nos. 21H05233 and 23H02052), the Core Research for Evolutional Science and Technology (CREST, JPMJCR24A5), Japan Science and Technology Agency (JST) and the World Premier International (WPI) Research Center Initiative, Ministry of Education, Culture, Sports, Science and Technology (MEXT), Japan.